# Verification and Validation of START: A Spent Nuclear Fuel Routing and Decision Support Tool

Caitlin Condon[a], Mark Abkowitz[b], Harish Gadey[a], Robert Claypool[c], Steven Maheras[a], Matthew Feldman[a], Erica Bickford[d]

[a]Pacific Northwest National Laboratory, 902 Battelle Blvd, Richland, WA 99354 (caitlin.condon@pnnl.gov)
[b]Vanderbilt University, 2201 West End Ave, Nashville, TN 37235
[c]Kanini Software Solutions Inc, 25 Century Blvd, Ste. 602, Nashville, Tennessee 37214, US
[d]U.S. Department of Energy (DOE), 1000 Independence Ave SW, Washington D.C., 20585

## INTRODUCTION

START – Stakeholder Tool for Assessing Radioactive Transportation – is a web-based, decision-support tool developed by the U.S. Department of Energy (DOE) to support the Office of Integrated Waste Management (IWM) [1].[a] Its purpose is to provide visualization and analysis of geospatial data relevant to planning and operating large-scale spent nuclear fuel (SNF) and high-level radioactive waste transport to storage and/or disposal facilities.

At present, the primary transport method for these shipments is expected to be via rail, operating predominantly on mainline track. For many shipment sites, however, access to this network will typically require initial use of a local/regional (short line) railroad or involve intermodal transport where the access leg is a movement performed by heavy-haul truck and/or barge. START has the ability to represent and analyze all of these transport options, with each transportation network segment containing site-specific physical and operational attributes. Of particular note are segment-specific accident rates and travel speeds, derived from recent data provided by the U.S. Department of Transportation, Bureau of Transportation Statistics, and other publicly available resources.

DOE anticipates that START users will include federal, State, Tribal and local government officials; nuclear utilities; transportation carriers; support contractors; citizen scientists; and other stakeholders. For this reason, START is designed to enable the user to represent a wide range of operating scenarios and performance objectives, with an emphasis on providing flexibility. In doing so, the tool makes extensive use of geographic information systems (GIS) technology for performing spatial analysis and map creation.

In addition to the transportation network itself, START includes a variety of data layers that characterize populations, land use and emergency response assets in proximity to potential shipment routes. Detailed attributes are provided for each point/link/polygon contained in each data layer that describe unique site-specific information such as population and transportation specific information. A suite of base map options is also provided for users to perform analyses and present results against the desired cartographic background.

Within this context, it is important to note that the population data used by START comes directly from LandScan [2] and is updated with successive LandScan releases; LandScan data sets are updated yearly. Both daytime and nighttime populations are incorporated in this tool. Utilizing the LandScan methodology represents a significant advancement over census population data because of the fidelity of LandScan that goes beyond Census tract and block aggregations. For this reason, START is able to more accurately calculate populations exposed to incident-free and dose-risk population exposure along transport routes. Moreover, it is not necessary to place population densities into bins (i.e., urban, suburban, rural) if a more precise estimate is desired.

Because several of START's land use databases describe national parks and forests, Tribal lands, and places of worship, among others, START also has the capability to measure impacts on cultural and environmentally pristine areas.

When assessing routing options and risk attributes, users may select a shipment origin and destination from predefined locations provided via a dropdown menu or may identify any desired shipment location within the contiguous U.S. via point-and-shoot functionality. When performing a routing analysis, the user may select from the following routing criteria in finding a route: 1) minimize travel time, 2) minimize route distance, 3) minimize population exposure, or 4) minimize a combination of travel time and population exposure, with ability to assign different importance weights assigned to each. START allows users to define these routing criteria but does not provide recommendations on which criteria to use for route generation. Once origin/destination, mode(s) and routing criteria are established, the user has the option to designate geographic locations to avoid as well as locations where the shipment is required to traverse; this is often used to accommodate where there may be shipment size and/or weight limitations, or where route restrictions are in place. Regarding the proximity of populations, land uses and emergency response assets to potential shipment routes, START considers two route buffer distances – 800 and 2,500 meters.

When a routing analysis is complete, START provides both summary (overall route) and detailed (route segment) performance measures and maps to support analysis and communication. Additionally, START allows users to report route analysis results according to political jurisdiction (i.e., by State, Tribal land, county, congressional district, military bases, State legislative districts). Collectively, these measures provide information on economic, safety, security, and environmental impacts.

Route analysis results can be exported from START in a variety of arrangements, including as pre-formatted reports, shapefile, comma-separated values (csv), and keyhole markup language (kml). This enables users to use START output in customized applications based on access to other in-house analysis and presentation tools. Users can also share routing analysis results with other users directly within the START platform.

Several other features are available in START to support user needs. Batch processing capability is available to situations where multiple routing analyses are being run concurrently (e.g., many shipment origins to a single shipment destination; many routing scenarios involving a specific shipment origin and destination). START can also accommodate photographic features, a convenient option when attempting to visualize potential obstacles along a candidate route. Additionally, measurement tools (e.g., area, distance, elevation, map coordinates) are available, as are smart mapping capabilities from which spatial information can be filtered or presented by theme.

Beyond performing routing analyses, START can be utilized for a number of other purposes, including 1) training preparations along routes (route proximity and response coverage provided by fire, police, hospitals and DOE Transportation Emergency Preparedness Program (TEPP) trained personnel), 2) communications (feasible transportation options connecting nuclear plants and DOE sites to line-haul networks), 3) environmental analyses (transportation dose estimates), and 4) integration with overall waste management systems analysis (inputs to DOE's Next Generation System Analysis Model, a backend systems analysis tool).

START development activities are ongoing, with work underway or anticipated in the following areas: 1) migrating the START application to the Amazon Web Services (AWS) Cloud, 2) developing enhanced EIS/NEPA analysis compatibility, 3) improving routing methodologies, and 4) considering addressing other complex waste types.

**Agile Software Development**

Potential scope was identified for improvements to better support the START user community and stakeholder requirements through a collaborative effort between the Pacific Northwest National Laboratory (PNNL) Verification and Validation (V&V) and the START development teams. An agile software development approach was employed by the teams in the interest of enhanced communication. Each round of software testing performed by the V&V team leads to feedback provided to the development team which results in an updated version of START. V&V tests are again performed on the latest release which leads to additional feedback. This continuous feedback iterative loop between the V&V and software development efforts is captured in Fig.1.

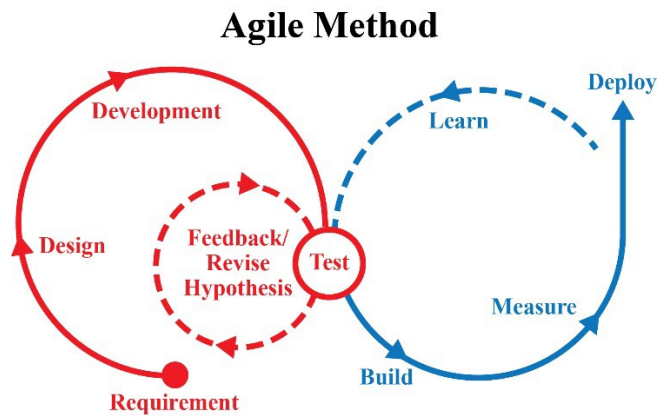

Fig. 1. Agile software development methodology

**START V&V EFFORTS**

The START V&V efforts for Fiscal Year (FY) 21 were carried out based on the official hypothetical test routes that were established in FY20 [3,4]. These included a total of 155 origin-destination pairs and a total of 310 routes (considering 2 buffer regions for each route). The following criteria were used in establishing the test routes:
- The destination was the geographical center of the continental United States.
- Test routes were created for all shutdown reactor sites, operating nuclear power plants, existing spent nuclear fuel storage facilities, and DOE facilities storing spent nuclear fuel or high-level waste.
- All routes were run with a minimum distance criterion.
- Each origin-destination pair had two independent routes: one representing each buffer zone at 800 and 2500 meters.
- All the criteria except the buffer zone were set identical between the two routes in an origin-destination pair.
- A unique testing number was assigned to each route in the interest of record keeping.
- As for mode of transportation, priority was given to rail only routes, followed by heavy haul truck to

rail, followed by heavy haul truck, and finally barge to rail routes.

Following each update to START, the V&V team recreated all the hypothetical test routes using the batch mode functionality in START. Routes are created using guidance from the official test series which helps in directly comparing results from two different releases or updates. Key attributes of the START outputs that were tested during FY21 include:
- Buffer zone population.
- Total route length.
- Population densities.
- Segment Population.
- Incident free dose.

Apart from this, the V&V team also discovered a few examples of route aberrations that will be discussed in this work.

**Buffer Zone Population**

Population numbers in buffer zones play an important role in impact evaluation and assessment. During FY21, START underwent an update that led to the release of START version 3.2.2. This provided an opportunity to observe population differences if any between the two START versions (3.2.1 vs 3.2.2). Relevant information was extracted for both the 800- and 2500-meters buffer zones from the LandScan raster data. Results from all 310 routes were compared and it was observed that the population difference between the two versions was under 1%. Although this difference is considerably small, the development team is currently investigating if this is originating because of variation in the route generation process between the two versions or some other factors within START 3.2.2.

After this, the second round of evaluation was conducted where data from START 3.2.2 was compared against results from ArcMap [5] and Quantum GIS (QGIS) [6], two popular GIS tools. It must be pointed out that both START, and ArcMap utilize Environmental Systems Research Institute (ESRI) software, however this effort was undertaken to observe and compare the results from two different tools that are using the same underlying software (i.e., and independent evaluation of the route using ArcMap compared to the START outputs that use ESRI tools). Data was extracted for both day and night using both buffer zones. The average population was computed by estimating a 2:1 weighted average of the night and daytime population respectively. A strong agreement was observed between the results from START and ArcMap since they were using the same underlying software from ESRI. Results from QGIS indicated that most of the population results were within 1% of each other with a select few routes exhibiting a difference within 5%.

**Total Route Length**

Total route length is one of the important outputs from a START run. Therefore, routes were initially created in START and the respective kml files were downloaded for independent analysis. For this study, QGIS was used as it provides a truly independent analysis of the results obtained from START. It was observed that differences in total route length was within 1% in most cases with a few cases resulting in a difference within 5%. Total route length was found to be consistent for the test routes when comparing START 3.2.1 and START 3.2.2.

**Population Densities**

Population densities play an important role for impact evaluation; therefore, it was included as a critical output in START. Differences were discovered between the estimates being produced by the V&V team and the START development team during FY20. Discussions with the development team shed light on the fact that the population densities in START were being calculated based on a summation of individual segment population densities. Further discussions with the START development team revealed that the best path forward might be to calculate the population densities for the entire route rather than summing the individual segments. This change in methodology addresses an important challenge of estimating population densities because of buffer zone overlaps between various segments. It is the expectation of the START development and the V&V team that opting for this methodology can help obtain population densities for various geographical regions such as an entire State or specific regions of a selected route and will provide a more useful analysis for the START users. This feature is currently under development by the START development team.

**Segment Population**

Segments in START are the individual building blocks for a route. Several hundreds or thousands of segments combine with each other to create a route. As segments are the fundamental components of a route, it was decided to study the population changes in the buffer zone at the segment level as well as the full route. A route with 96 segments was selected for this exercise and the kml file was downloaded from START. QGIS was used to dissect the kml file into various segments since segment level data download is currently not available in START. These kml files were subsequently loaded in QGIS to evaluate the population within each segment. Again, QGIS was chosen since its framework was completely independent of ESRI software. The results indicated that most of the segments had populations within 1% of each other, while there were a few segments where a difference above 5% was observed, there was only one segment where a buffer population difference of 12.05% was recorded.

**Incident Free Dose**

In START version 3.2.2, the incident free dose at the segment level was being provided. Dose to both the public as well as the crew were available in this version. These dose calculations were performed based on a concept of unit dose factors (UDFs) that was introduced by Connolly in 2015. The gamma UDFs were developed using a point source model while the neutron UDFs were modeled using a discrete ordinance neutron model. The START development team aided with providing the pseudocode for implementing the UDF methodology. It is worth mentioning that the UDFs are currently only established for the 800-meter buffer zone and dose calculations were implemented in START by assuming that 89% of the dose is originating from gammas while 11% of the dose is contributed by neutrons. An assumption was also made that packages were at a regulatory limit of 1mSv/h at a distance of 2 meters. Based on the equations from the Connolly report and the START development team's implementation of this methodology, the V&V team performed calculations to verify that there are no significant differences between the results from START and the confirmatory calculations performed by the V&V team. Based on the V&V efforts, the following conclusions were reached:

- The dose to the public (on-link and off-link) and the crew matches the results from START for all modes of transportation (barge, rail, and heavy haul truck).
- The START tool does not report values below $1 \times 10^{-8}$ mSv and any values above $5 \times 10^{-9}$ mSv but below $1 \times 10^{-8}$ mSv are simply reported as $1 \times 10^{-8}$ mSv.
- Values below $5 \times 10^{-9}$ mSv are reported as 0 mSv.
- The V&V effort concluded that any differences in dose values between the START outputs and the confirmatory calculations by the V&V team were under $5 \times 10^{-9}$ mSv.

**Route Aberrations**

Apart from these key outputs from START, several studies were also carried out to investigate challenges related to route aberrations. These include:
- Variations between Shape and kml files.
- Unusual deviations and route jumps from designated paths.
- Variable origin location with respect to actual source location.
- Unused and unknown mode of transportation.

Fig. 2 shows an example where there is a deviation between the downloadable shape file and kml file for a route.

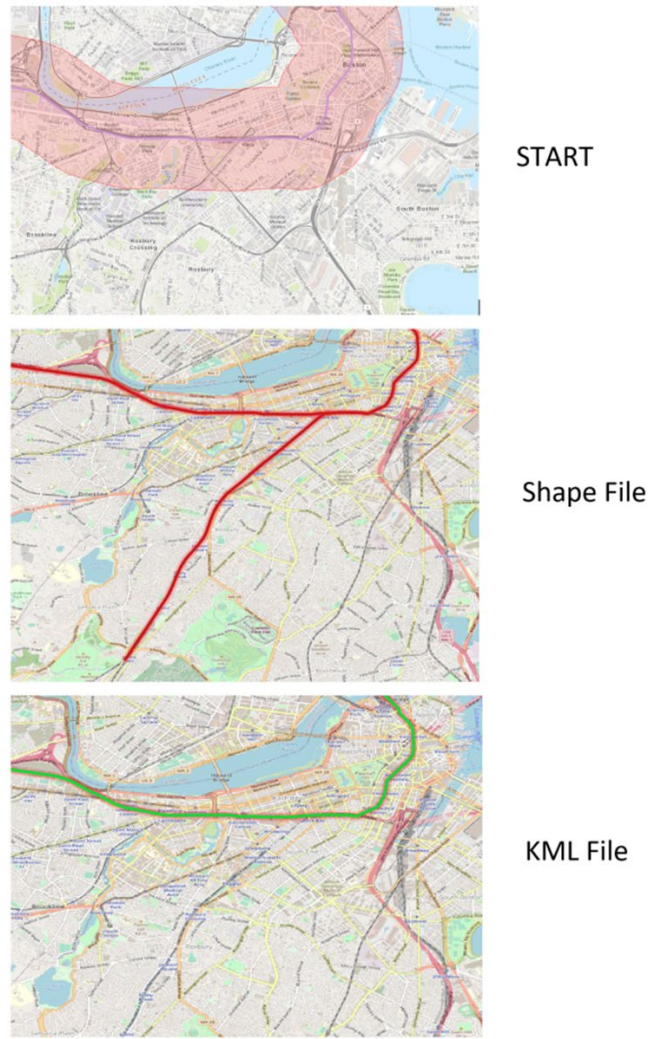

Fig. 2. Examples of START Route Shape File Aberration

**CONCLUSIONS**

The V&V efforts for START version 3.2.2 found good agreement between START results from 3.2.1 and the independent assessments for full route buffer zone population and route distance. The V&V utilized QGIS as a fully independent GIS tool and ArcMap, which is not independent from START but provides confirmation that START outputs can be replicated in other ESRI products. In all test cases, the route buffer zone population and route distance showed good agreement between START and the independent results. The percent difference was less than +/- 5% in all cases, with the majority of cases less than +/- 1%. Analysis of population and dose information available at the segment level of a route also showed good agreement between START version 3.2.2 and independent analysis. The START V&V team as well as the START development team will continue collaborative efforts to identify areas of potential improvement within START. PNNL-SA-174406.

**ENDNOTES**

[a] This is a technical paper that does not take into account contractual limitations or obligations under the Standard Contract for Disposal of Spent Nuclear Fuel and/or High-Level Radioactive Waste (Standard Contract) (10 CFR Part 961). For example, under the provisions of the Standard Contract, spent nuclear fuel in multi-assembly canisters is not an acceptable waste form, absent a mutually agreed to contract amendment.

To the extent discussions or recommendations in this paper conflict with the provisions of the Standard Contract, the Standard Contract governs the obligations of the parties, and this paper in no manner supersedes, overrides, or amends the Standard Contract.

This paper reflects technical work which could support future decision making by the US Department of Energy (DOE or Department). No inferences should be drawn from this paper regarding future actions by DOE, which are limited both by the terms of the Standard Contract and Congressional appropriations for the Department to fulfil its obligations under the Nuclear Waste Policy Act including licensing and construction of a spent nuclear fuel repository.

**ACKNOWLEDGEMENTS**

Pacific Northwest National Laboratory is operated by Battelle Memorial Institute for the US Department of Energy under Contract No. DE-AC05-76RL01830. This work was supported by the US Department of Energy Office of Integrated Waste Management.